\documentclass[conference]{IEEEtran}
\IEEEoverridecommandlockouts
\usepackage{cite}
\usepackage{amsmath,amssymb,amsfonts}
\usepackage{algorithmic}
\usepackage{graphicx}
\usepackage{todonotes}
\usepackage{textcomp}
\usepackage{xcolor,ulem}
\usepackage[]{algorithm2e}

\def\BibTeX{{\rm B\kern-.05em{\sc i\kern-.025em b}\kern-.08em
    T\kern-.1667em\lower.7ex\hbox{E}\kern-.125emX}}
\begin{document}

\title{Cellular Coverage-Aware Path Planning for UAVs}

\author{
\IEEEauthorblockN{Sibren De Bast}
\IEEEauthorblockA{\textit{ESAT - TELEMIC} \\
\textit{KU Leuven}\\
Leuven, Belgium \\
sibren.debast@kuleuven.be}
\and
\IEEEauthorblockN{Evgenii Vinogradov}
\IEEEauthorblockA{\textit{ESAT - TELEMIC} \\
\textit{KU Leuven}\\
Leuven, Belgium \\
evgenii.vinogradov@kuleuven.be}
\and
\IEEEauthorblockN{Sofie Pollin}
\IEEEauthorblockA{\textit{ESAT - TELEMIC} \\
\textit{KU Leuven}\\
Leuven, Belgium \\
sofie.pollin@kuleuven.be}
}

\maketitle

\begin{abstract}
Up until now, path planning for unmanned aerial vehicles (UAVs) has mainly been focused on the optimisation towards energy efficiency. However, to operate UAVs safely, wireless coverage is of utmost importance. Currently, deployed cellular networks often exhibit an inadequate performance for aerial users due to high amounts of intercell interference. Furthermore, taking the never-ending trend of densification into account, the level of interference experienced by UAVs will only increase in the future. For the purpose of UAV trajectory planning, wireless coverage should be taken into account to mitigate interference and to lower the risk of dangerous connectivity outages. In this paper, several path planning strategies are proposed and evaluated to optimise wireless coverage for UAVs. A simulator using a real-life 3D map is used to evaluate the proposed algorithms for both 4G and 5G scenarios. We show that the proposed Coverage-Aware A* algorithm, which alters the UAV's flying altitude, is able to improve the mean SINR by 3-4dB and lower the cellular outage probability by a factor of 10. Furthermore, the outages that still occur have a 60\% shorter length, hence posing a lower risk to induce harmful accidents.
\end{abstract}


\section{Introduction}
\label{sec:intro}

In recent years, unmanned aerial vehicles (UAVs) have sparked a lot of research driven by the promise of many different exciting applications. Most of these applications expect the UAV to be a user of a telecommunications network. For instance, UAVs used for search-and-rescue during avalanches~\cite{Silvagni17} are in need of a low-latency wireless link in order to successfully locate survivors, while surveillance UAVs cannot deliver reliable camera footage without a high-throughput link. For more applications and a comprehensive understanding of UAV communications, \cite{BlueSky} can be consulted.

Although wireless links are of utmost importance in UAV applications, modern cellular networks are unable to deliver a reliable connectivity due to intercell interference\cite{LTEsky}. This intercell interference is caused by the propagation-favourable line-of-sight (LoS) links between the UAV and nearby Base Stations (BS). When a UAV gains altitude, the number of BSs in sight will rise, causing extra interference, which lowers the quality of the network service.

This problem can be solved in two ways, the positions of the deployed BSs can be adjusted, or the UAVs' trajectory can be altered. Since changing the deployed infrastructure is hard, due to local legislation and high costs, we assume that the it is fixed. Therefore, the focus of this paper will be on controlling the trajectory of the UAV to optimise its wireless coverage. This study considers how coverage-aware path planning influences the wireless coverage of UAVs, both for current 4G and future 5G networks.

In the past, path planning research for UAVs was mainly focused on optimisation towards energy efficient object-avoidance\cite{Goerzen10}. However, recently, some studies have been published that optimise the trajectory for UAVs considering energy and wireless communication performance. Zeng and Zhang propose a trajectory optimisation method that maximises the number of bits sent, normalised with the energy needed to propel the UAV \cite{Zeng17}. Mardani et al. propose A*-based algorithms to optimise the trajectory of a UAV to maximise its video stream quality \cite{Mardani18}. However, both of them do not consider object avoidance, moreover, they use simple wireless models to evaluate the proposed methods. Therefore, considering the interference rich urban environments of real cities, it is impossible to guarantee a low chance of wireless outages when using current state-of-the-art techniques. 


%
In this paper, different path planning algorithms are proposed and analysed. We evaluate these methods based on signal-to-interference-and-noise ratio (SINR), outage probability and outage duration. To evaluate them, a realistic urban 3D simulator is used and three important frequency bands are assessed using their corresponding channel models. 

This paper is organised as follows. Section \ref{sec:system} delineates the used system model and the considered path planning methods. The performance evaluation of these methods can be found in Section \ref{sec:evaluation}. Finally, Section \ref{sec:conclusion} concludes this study.

\section{System model}
\label{sec:system}
In this section, the used system model is explained. First, the virtual environment used to evaluate the performance is described. Next, we evaluate the simulator to calculate the coverage in this virtual environment. Afterwards, the constraints on the UAVs' movements are defined and finally, the proposed path planning methods are described.

\subsection{Virtual Environment}
\label{sec:env}
Flying UAVs in real-life takes a lot of time and planning, therefore a virtual environment to fly UAVs is preferred.
For this purpose, we used a freely available 3D map of Flanders, Belgium \cite{geopunt}. The map consists of height data (with centimetre precision) generated by a Lidar and has a latitude and longitude resolution of 1~m by 1~m. Each pixel contains the height of the surface, which allows us to generate a 3D environment to use as a test scenario.

\begin{figure}[!ht]
  \centering
  \includegraphics[width=\linewidth]{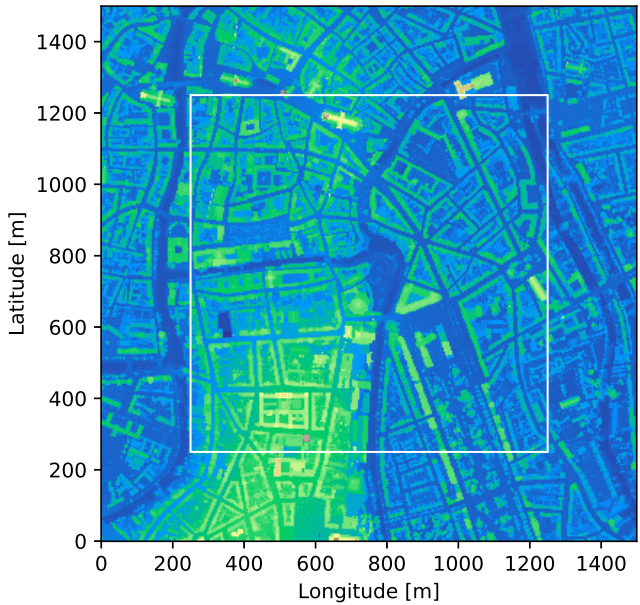}
  \caption{The selected region of interest for the evaluation of the path planning strategies. The map has a size of 1.5 km by 1.5 km, while the virtual UAVs fly within the 1km$^2$ white box. The map has a resolution of 1m and the colour value corresponds to the height of the surface.}
  \label{fig:dsm}
\end{figure}

Since Ghent represents a middle-sized European city very well, a 1.5 km by 1.5 km area covering this city was selected. Fig. \ref{fig:dsm} shows the selected region of the map. Next, real-life BS locations are imported on the map. Nineteen BSs of the Belgian mobile provider ``Proximus'' are located in the considered area. These locations are provided by the Belgian Institute for Postal services and Telecommunications (BIPT)\cite{bipt}.

\subsection{Coverage Simulator}

The 3D radio coverage simulator presented by Colpaert et al.~\cite{Colpaert18} is used in this study to estimate the altitude-dependent SINR. The simulator i) uses the 3D environment map to calculate deterministic propagation conditions (Line-of-Sight (LoS) and Non-Line-of-Sight (NLoS)) and ii) calculates the signal and interference levels depending on the conditions, link distances, and UAV height. The previous version of the simulator was extended by adding support for the 3.5~GHz \cite{3gpp} and 28~GHz \cite{rappaport15} frequency bands, since they are chosen as the pioneering 5G bands in Europe.

Since BSs have relatively large coverage areas (especially in the air, due to the favourable propagation conditions), the border of the map will also be influenced by BSs outside the region of interest. Therefore, only the central 1~km$^2$ area will be evaluated. The white square on Fig. \ref{fig:dsm} denotes this area. Next, the simulator calculates the received power from each BS at each position in this area, up to 120~m above the surface. For each location, the BS with the highest received power is chosen as the serving BS, allowing for the calculation of the SINR for each location in this 3D space. The obtained SINR values represent the worst case scenario, since the calculations assume that all interfering BSs are under full load, hence transmitting continuously.

To evaluate the performance for both 4G and 5G networks, the path planning methods have to be evaluated for both sub-6~GHz and mmWave frequencies. In the case of 4G, the 1.8~GHz band is  used by ``Proximus'' in Ghent, therefore, this frequency will be evaluated. The bandwidth used to calculate the SINR for 4G is 20~MHz, the maximum bandwidth of LTE. In the case of 5G in Europe, the 3.5~GHz and 28~GHz bands are targeted. Hence, these frequencies are considered in the simulations, using a bandwidth of 100~MHz and 400~MHz respectively, which are the maximum bandwidths for New Radio (NR) in these bands.

\subsection{UAV movements}
\begin{figure}[!ht]
\centering
  \includegraphics[width = \linewidth]{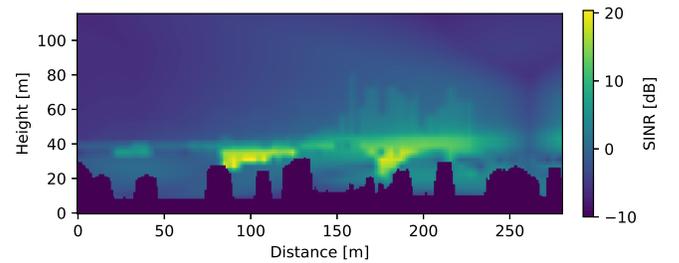}
  \caption{A path profile of the 3D coverage map. At the bottom, buildings and the terrain can be seen, while the sky is coloured following the SINR, a lighter colour means a higher SINR and vice versa.}
  \label{fig:intersection}
\end{figure}

When flying from one point to another, a straight line is the shortest path. However, for safety reasons, a UAV must avoid outages and ensure an acceptable level of the command and control link's reliability and quality. For this purpose, the height of the drone is controlled while flying between two points. Hence an image can be generated, showing the path profile in which the height of the drone can be altered. Fig.~\ref{fig:intersection} shows an example of the path profile, where at the  bottom the buildings and the terrain can be seen in a dark colour, while the sky is coloured in accordance to the SINR.

Since the virtual environment has a resolution of 1m by 1m, we define that all possible UAV positions lay on a grid of 1m by 1m. Furthermore, we define that the drone can move vertically, horizontally, and diagonally to adjacent positions. As a result, moving horizontally or vertically will add 1m to the path length, while moving diagonally adds 1.41m.

For safety purposes, UAVs are not allowed to fly close to obstacles. Therefore, a minimum height above the terrain and a guard distance between the UAV and buildings are defined, measuring 10~m and 3~m, respectively. This gives enough clearance between the obstacles and the UAV to allow for a safe operation.

\subsection{Path planning}
In this subsection, we delineate algorithms for UAV path planning. First, some very basic and rudimentary algorithms are presented. These methods set a baseline to compare the more advanced algorithms, which will be presented at the end.

\subsubsection{Straight path}
The first path planning method, likely the most widely used one, that will be evaluated is a simple straight path. The UAV rises to a height high enough to clear all terrain and buildings on the straight line to the destination and flies on this trajectory.

\subsubsection{22m Above Ground Level (AGL)}
As shown by Colpaert et al.\cite{Colpaert18}, the mean SINR above the city of Ghent is the highest while flying at an altitude of 22~m AGL. Therefore, when using this method, the height of the UAV will be altered to stay 22~m above the terrain. When a building higher than 19~m is encountered, the height of the UAV is altered to ensure a clearance of 3~m between the building and the UAV.

\subsubsection{Optimal Coverage Height (OCH)}
This method uses the generated coverage maps by adjusting the height of the UAV to the height with the highest SINR for each step in the horizontal direction of the trajectory. The advantage of this method is that the height is continuously adjusted to the optimal height. The drawback is that this method can add a lot of length to the path every time the optimal height changes.

\subsubsection{Coverage-Aware A* (CAA*)}
A* path planning, proposed by Hart et al. \cite{hart68}, is based on the Dijkstra algorithm, but it adds an extra heuristic to the cost of the nodes to give a sense of direction. A commonly used heuristic cost is the euclidean distance to the goal. We can put this as:
$$f(n) = g(n) + h(n)$$
Where $n$ is the next node on the path, $g(n)$ is the cost to go from the starting node to node $n$ and $h(n)$ is the heuristic cost to go from $n$ to the destination. Here a node is a pixel on the path profile as seen in Fig. \ref{fig:intersection}.

A* path planning minimises $f(n)$ to find the shortest path. First, the starting node is selected as the current best node. Next, the neighbouring nodes to the current best node are evaluated and added to the priority queue. The previous best node is removed from this queue. Afterwards, the priority queue is sorted, based on the $f$-value of its nodes. The node in the queue with the lowest $f$-value, becomes the new best node. This process is repeated until the goal node is reached. All nodes keep track of its predecessor, in this way the path can be reconstructed by repetitively selecting the preceding node and adding it to the path.

This algorithm was altered to find a short path while avoiding regions with a low coverage, by adding a coverage-dependant cost $c(n)$ to $g(n)$. This cost is high when the trajectory goes along an area with a low SINR and low during high SINR periods. It is calculated in the folling way:
$$c(n) = \frac{SINR_{threshold} - SINR_n}{c_{normalisation}}$$
where $SINR_{threshold}$ is a defined threshold that decides upon if the cost will add a penalty or a reward. $SINR_n$ is the SINR at the location of node $n$ and $c_{normalisation}$ is a normalisation factor to influence the size of the cost. Their values have been numerically determined as -3 and 1.5, respectively.

\section{Performance evaluation}
\label{sec:evaluation}
In this section, the performance of the proposed path planning algorithms is evaluated. For this purpose, 3000 random pairs of starting and destination points in Ghent were generated. Each of the algorithms calculated a path between the two locations, for each of the 3000 trajectories. Along these trajectories, the SINR was logged, which will be assessed in this section. First, the achieved SINR along the paths is evaluated, followed by the outage probability and duration. To end, the length of the different trajectories is compared.

\subsection{SINR}
To evaluate the SINR for each of the three scenarios, the Cumulative Density Function (CDF) of the SINR of the proposed methods is calculated. In addition, the mean SINR is calculated and reported in Table \ref{tab:sinr}.

\begin{table}[!ht]
\centering
\caption{The mean SINR of the evaluated path planning methods in the three different scenarios.}
\begin{tabular}{| l||c|c|c| }
 \hline
                           & 4G     & 5G (3.5~GHz) & 5G (28~GHz) \\
 \hline
 \hline
  Straight path            & 1.61dB & 1.62dB      & -0.12dB \\
  22m AGL                  & 3.08dB & 3.12dB      & 1.29dB \\
  Coverage-Aware A*        & 4.88dB & 4.97dB      & 4.43dB \\
  Optimal coverage height  & 5.46dB & 5.58dB      & 4.80dB \\
 \hline
\end{tabular}
\label{tab:sinr}
\end{table}

\begin{figure}[!ht]
\includegraphics[width =0.9\linewidth]{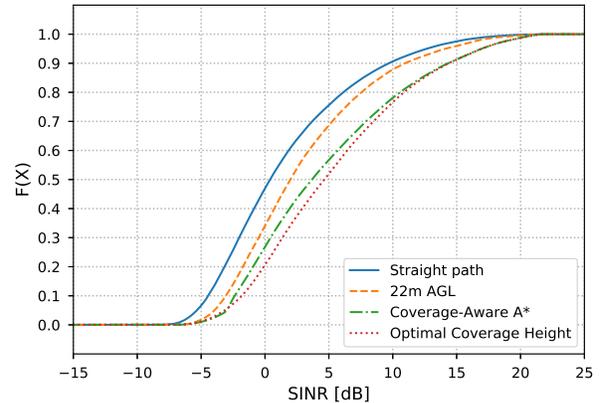}
\caption{The SINR CDF of 4G.}
\label{fig:cdf_LTE_20}
\end{figure}

Fig. \ref{fig:cdf_LTE_20} depicts the CDF for 4G. It can clearly be seen that including wireless coverage information into the path planning method improves the mean SINR. The 22~m-AGL method improves the mean SINR level by 1.4~dB. Moreover, CAA* and OCH improve the mean SINR level by 3.26~dB and 3.79~dB respectively. Although OCH has the highest gain, CAA* has a small benefit when operating in the low-SINR range.


\begin{figure}[!ht]
\includegraphics[width =0.9\linewidth]{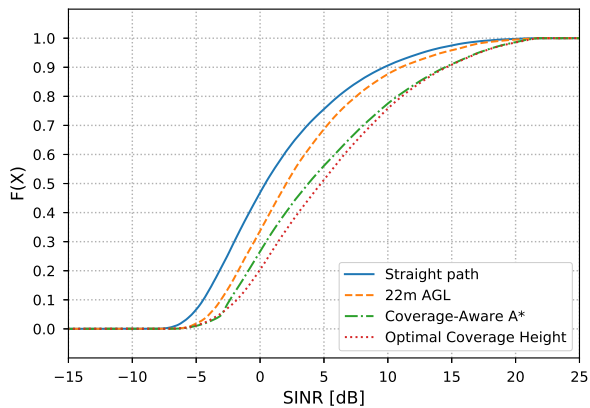}
\caption{CDF of the SINR of sub-6~GHz 5G.}
\label{fig:cdf_nr_100}
\end{figure}


\begin{figure}[!ht]
\includegraphics[width =0.9\linewidth]{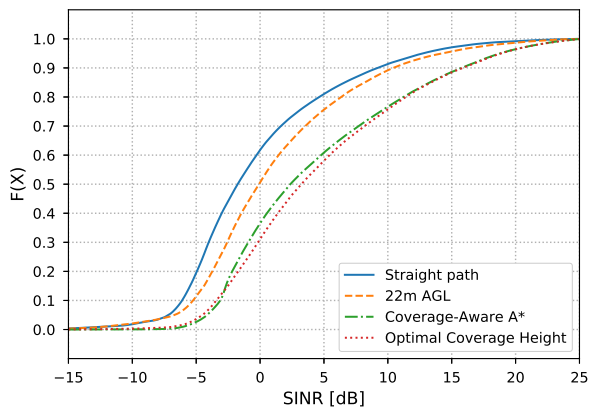}
\caption{CDF of the SINR of mmWave 5G.}
\label{fig:cdf_28_400}
\end{figure}

For sub-6~GHz 5G, the results can be found in Fig. \ref{fig:cdf_nr_100}. Here, the same conclusions are true as for 4G, this is due to the interference limited channel. As can be seen in Fig. \ref{fig:cdf_28_400}, the situation is different for mmWave 5G. The average SINR at these high frequencies is lower. However, by using a coverage-aware path planning method, an almost equal performance can be achieved in comparison to sub-6~GHz systems. At these mmWave frequencies, CAA* and OCH improve the mean SINR level by 4.55~dB and 4.92~dB respectively.

\begin{figure}[!ht]
  \includegraphics[width = \linewidth]{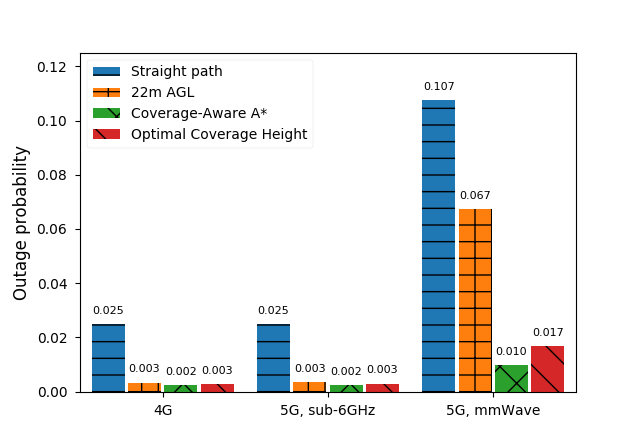}
  \caption{The outage probability for the different path planning methods. In all cases CAA* has the best performance of the evaluated methods, it reduces the outage probability with a factor of 10 in comparison to flying in a straight line.}
  \label{fig:outageprobability}
\end{figure}

\subsection{Outages}
An outage is defined as a situation when the SINR~$\leq$~$-$6~dB. This is the lowest possible SINR for LTE (4G) to have a basic wireless connection. Note that this SINR-level will not support video streaming, although, it will support a basic control link. This basic control link is essential for a save operation of UAVs. To evaluate the performance in terms of outages, both the outage probability and duration are discussed.

Fig. \ref{fig:outageprobability} illustrates the outage probability. Of all methods, CAA* performs the best and lowers the outage probability to 0.2\% for sub-6~GHz bands and 1.0\% for 28~GHz bands. This is a reduction by a factor 10 in comparison with flying in a straight line. 

\begin{figure}[!ht]
\centering
  \includegraphics[width = \linewidth]{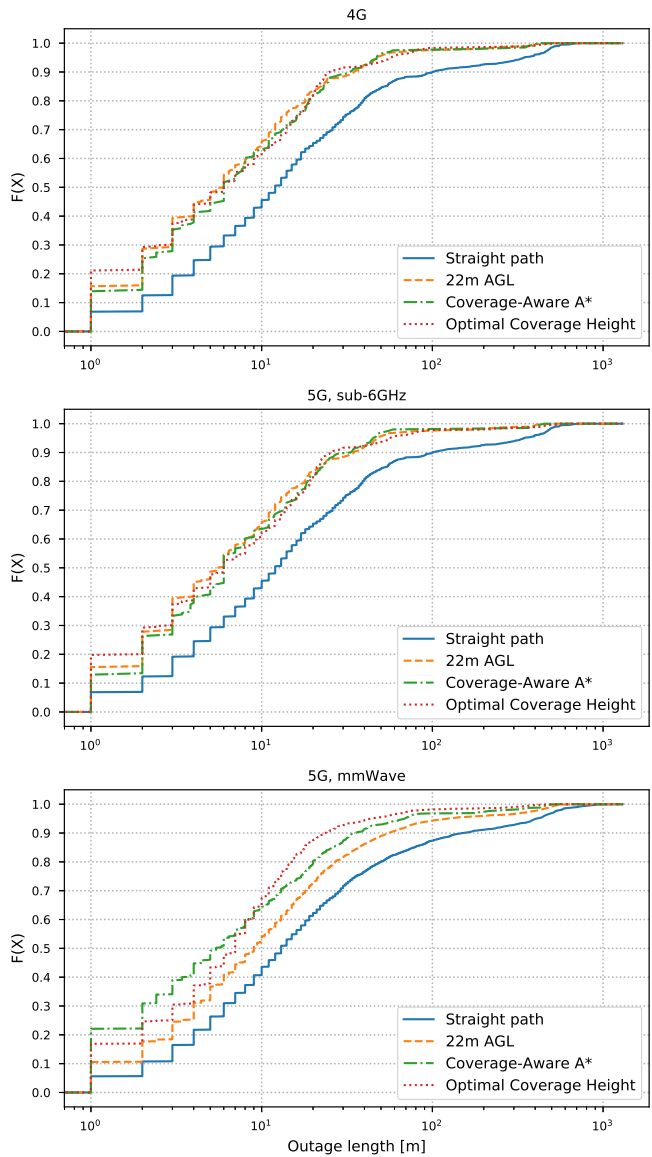}
  \caption{The CDFs of outage duration for the different path planning methods. They show the spread of the duration of the outages. In all scenarios, OCH has the shortest outages and the Straight path method has the longest.}
  \label{fig:outageduration}
\end{figure}

To see the full picture, the duration of the encountered outages is from utmost importance, as small outages can be dealt with, while long outages prohibit a UAV from flying. To assess this, the mean outage duration is presented in Table \ref{tab:outagelength} and the outage duration CDF is shown in Fig. \ref{fig:outageduration}. In all three scenarios, the outage duration may be greatly reduced using coverage-aware path planning. Moreover, CCA* and OCH show a consistent improvement in all three scenarios, resulting in the outage durations being shortened with $\geq$ 60\%.


\begin{table}[!ht]
\centering
\caption{The mean outage duration of the evaluated path planning methods in the three different scenarios.}
\begin{tabular}{| l||c|c|c| }
 \hline
                           & 4G     & 5G (3.5~GHz) & 5G (28~GHz) \\
 \hline
 \hline
  Straight path            & 49.9 m & 50.0 m      & 61.9 m \\
  22m AGL                  & 18.2 m & 18.2 m      & 32.7 m \\
  Coverage-Aware A*        & 19.0 m & 18.6 m      & 21.8 m \\
  Optimal coverage height  & 18.1 m & 20.1 m      & 15.5 m \\
 \hline
\end{tabular}
\label{tab:outagelength}
\end{table}

\subsection{Path length}
While planning UAV trajectories, improving SINR is a desirable objective. However, since UAVs are energy constraint, SINR improvement may not come at the cost of a very high length of the trajectory. Therefore, the path length of the proposed path planning methods is evaluated. To compare the path lengths, the 3000 trajectories generated are first averaged and afterwards normalised using the shortest possible path length, a straight line. 

\begin{figure}[!htb]
\centering
  \includegraphics[width = \linewidth]{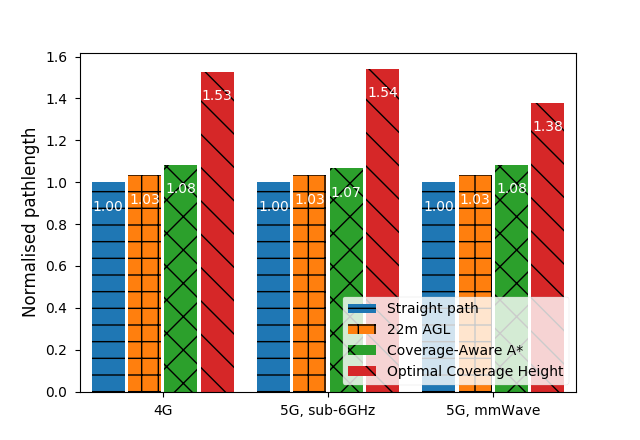}
  \caption{The normalised path length for the different path planning methods. CAA* adds on average 7-8\% to the path length, for OCH, this is 38-54\%.}
  \label{fig:pathlength}
\end{figure}

Fig. \ref{fig:pathlength} illustrates how the different methods affect the average path length. It shows how OCH adds 38-54\% to the total path length, which is unacceptable for UAV path planning. However, CAA* only adds 7-8\% to the total path length, while still improving the SINR almost as much as OCH. As a result, CCA* is a more viable option than OCH.

\section{Conclusions}
\label{sec:conclusion}
This study examined the effect of making UAV path-planning methods wireless-coverage aware. The past literature mainly focused on energy-efficient path planning for UAVs, without assessing the cellular coverage along the trajectory. The scarce literature that did include the optimisation of wireless connectivity, used too simplistic scenarios and propagation models. Moreover, object avoidance was not considered in these studies. 
Our analyses are supported by realistic simulations using state-of-the-art 3D channel models. Furthermore, they revealed that by altering the height of the drone along a predefined trajectory, the proposed Coverage-Aware A* path planning algorithm enhances the mean experienced SINR by 3.31 dB at sub-6~GHz frequencies and 4.55dB at 28~GHz. Moreover, it reduces the probability of a wireless outage by a factor of 10. In addition, the mean duration of the experienced outages is reduced by $\geq$~60\%, further improving the reliability of the UAVs' cellular connection. All of these benefits are reached while only increasing the path length by 7-8\%.

\section{Acknowledgements}
This research was partially funded by the Research Foundation Flanders (FWO), project no. S003817N (OmniDrone), and by a Research Foundation Flanders (FWO) SB PhD fellowship, grant no. 1SA1619N (Sibren De Bast).


\begin{thebibliography}{00}



  \bibitem{Silvagni17}
  M. Silvagni, A. Tonoli, E. Zenerino, and M. Chiaberge.
  ``Multipurpose UAV for search and rescue operations in mountain avalanche events.''
  Geomatics, Natural Hazards and Risk, 8(1):18–33, 2017.

  \bibitem{LTEsky}
  B. V. Der Bergh, A. Chiumento, and S. Pollin,
  ``LTE in the sky: trading off propagation benefits with interference costs for aerial nodes'',
  IEEE Communications Magazine, May 2016.
  
  \bibitem{BlueSky}
  E. Vinogradov, H. Sallouha, S. De Bast, M. M. Azari and S. Pollin,
  "Tutorial on UAVs: A Blue Sky View on Wireless Communication",
  Journal of Mobile Multimedia; Vol. 14; iss. 4; pp. 395 - 395, 2018
  
  \bibitem{Goerzen10}
  C. Goerzen, Z. Kong, and B. Mettler,
  ``A Survey of Motion Planning Algorithms from the Perspective of Autonomous UAV Guidance'',
  Journal of Intelligent Robot Systems, 57: 65, 2010
  
  

  \bibitem{Zeng17}
  Y. Zeng, and R. Zhang
  ``Energy-Efficient UAV Communication With Trajectory Optimization,''
  IEEE Transactions on Wireless Communications, vol. 16, no. 6, pp. 3747-3760, June 2017
  
  \bibitem{Mardani18}
  A. Mardani, M. Chiaberge, and P. Giaccone, 
  ``Communication-Aware UAV Path Planning,''
  IEEE International Conference on Wireless for Space and Extreme Environments (WiSEE), 2018

  \bibitem{geopunt}
  Geopunt, ``DHMV II'',\\
  http://www.geopunt.be/actualiteit/2016/april/dhmvii-volledig
  
  
  \bibitem{bipt}
  BIPT, ``Antennas: cartography'',\\
  https://www.bipt.be/en/operators/radio/antennas-site-sharing/antennas-cartography
  
  \bibitem{Colpaert18}
  A. Colpaert, E. Vinogradov, and S. Pollin;
  ``Aerial Coverage Analysis of Cellular Systems at LTE and mmWave Frequencies Using 3D City Models.'',
  Sensors 2018, 18, 4311, December 2018.
  
  \bibitem{3gpp}
  3GPP, ``Enhanced LTE Support for Aerial Vehicles; Technical Report 36.777'', Sophia Antipolis, France, 2017.

  \bibitem{rappaport15}
  G. MacCartney, T. Rappaport, M. Samimi, and S. Sun
   ``Wideband Millimeter-Wave Propagation Measurements and Channel Models for Future Wireless Communication System Design.''
  IEEE Trans. Commun. 2015, 63, 3029–3056.
  
  \bibitem{hart68} 
  P. E. Hart, N. J. Nilsson, and B. Raphael, 
  "A Formal Basis for the Heuristic Determination of Minimum Cost Paths," in IEEE Transactions on Systems Science and Cybernetics, vol. 4, no. 2, pp. 100-107, July 1968.

\end{thebibliography}
\end{document}